\documentclass[reprint, superscriptaddress, secnumarabic, amssymb, nobibnotes, aps, prl]{revtex4-1}
\usepackage{graphicx}
\usepackage{epstopdf}
\usepackage[T1]{fontenc}
\usepackage[latin9]{inputenc}
\usepackage{amsbsy}
\usepackage{gensymb}

\begin{document}

\title{\textrm{Superconducting Properties of the Non-Centrosymmetric Superconductor Re$_{6}$Hf}}
\author{D. Singh}
\affiliation{Indian Institute of Science Education and Research Bhopal, Bhopal, 462066, India}
\author{A. D. Hillier}
\affiliation{ISIS facility, STFC Rutherford Appleton Laboratory, Harwell Science and Innovation Campus, Oxfordshire, OX11 0QX, UK}
\author{A.~Thamizhavel}
\affiliation{Department of Condensed Matter Physics and Materials Science, Tata Institute of Fundamental Research, Mumbai 400005, India}
\author{R. P. Singh}
\email[]{rpsingh@iiserb.ac.in}
\affiliation{Indian Institute of Science Education and Research Bhopal, Bhopal, 462066, India}

\date{\today}
\begin{abstract}
\begin{flushleft}

\end{flushleft}
The synthesis and detailed characterization of the noncentrosymmetric superconductor Re$_{6}$Hf using powder x-ray diffraction (XRD), magnetization, transport and thermodynamic measurements are reported. XRD confirms the crystal structure of Re$_{6}$Hf to be cubic noncentrosymmetric $\alpha$-\textit{Mn} with a lattice parameter a = 9.6850(3)\text{\AA}. Resistivity, DC and AC magnetization confirms the bulk superconducting transition T$_{c}^{onset} \sim $ ~ 5.96 K. The lower critical field H$_{C1}(0)\approx$ 5.6  mT  and upper critical H$_{C2}(0)$ was found to be around 12.2 T which is greater than the Pauli limiting field, indicating unconventional behavior in Re$_{6}$Hf and leads to the possibility of triplet pairing. A sharp discontinuity in the specific heat data was observed around T$_{c}$ indicating bulk superconductivity, and the normalized specific heat jump is $\Delta C_{el}$/$\gamma_{n}$T$_{c}$ = 1.53  which is in close agreement with BCS value. Superconducting electronic specific heat data is fitted reasonably well for an exponential relation with Sommerfeld coefficient ($\gamma$) showing a linear relation with magnetic field suggesting that in Re$_{6}$Hf dominant pairing channel is s-wave with isotropic superconducting gap parameter without nodes. 
\end{abstract}

\keywords{ }

\maketitle

\section{INTRODUCTION}

In 1957, Bardeen, Cooper and Schrieffer (BCS) explained the origin of superconductivity using concept of Cooper pairs \cite{BCS}. Cooper pairs are constructed from electrons that have equal and opposite crystal momenta and opposite spin. The pairs are formed via an electron- phonon interaction. Because electrons are fermions, the wavefunction must be antisymmetric  with  respect to the exchange of the two electrons. The two electron wave function can be made asymmetric by considering the spatial and spin components. If spatial wavefunction is symmetric/anti-symmetric, the spin wavefunction must be anti-symmetric/symmetric. This is termed as spin- singlet state /spin-triplet state. This scenario is only possible if inversion symmetry is present in crystal structure \cite{ander1, ander2}.\\

There has been an intense theoretical and experimental studies on superconducting system lacking spatial inversion symmetry, after the discovery of unconventional superconductivity in noncentrosymmetric heavy fermion compound CePt${_3}$Si \cite{Bauer2004}. In NCS's absence of center of inversion induces Rashba-type antisymmetric spin orbit coupling (ASOC) due to an asymmetric potential gradient along the crystal axis; this momentum dependent spin structure in consequence lifts the degeneracy of
the conduction band electrons at Fermi surface resulting into the possible mixing of spin-singlet and spin-triplet
pair states \cite{vm,rashba,kv,pa,fujimoto}. The magnitude of ASOC splitting, actually dictates the extent of mixing of spin singlet and spin-triplet components. The possible mixing of pair states in noncentrosymmetric superconductors may lead to various novel superconducting properties such as upper critical field exceeding Pauli limit, line or point nodes in superconducting gap parameter resulting into nodal excitations in penetration depth, heat capacity and other thermodynamic measurements, time reversal symmetry breaking in superconducting condensate.\\

Apart from CePt${_3}$Si \cite{Bauer2004} several other Ce and U-based heavy fermionic NCS's CeRhSi${_3}$ \cite{kimura2005, kimura2007}, CeIrSi${_3}$ \cite{cis1,cis2}, CeCoGe$_{3}$ \cite{ccog}, UIr \cite{UIr}, UGe$_{2}$ \cite{Uge} were discovered with applied  pressure, but it doesn't resolve the problem as strong electron correlations in this f-electron systems with magnetic quantum criticality hinder the pursuit to realize the effects of ASOC and absence of inversion center on superconductivity. Superconductivity in these systems is expected to be due to non-conventional pairing mechanism, most likely driven by magnetic fluctuations.\\

Several weakly and strongly correlated non-centrosymmetric superconducting systems e.g. Li$_{2}${(Pd,Pt)}$_{3}$B \cite{lpt1,lpt2,lpt3,lpt4}, LaNiC$_{2}$ \cite{lnc1,lnc2}, Re$_{3}$W \cite{rw}, Mg$_{10}$Ir$_{19}$B$_{16}$ \cite{mib}, La $\textit{M}$$\textit{P}$($\textit{M}$=Ir, Rh and $\textit{P}$=P, As) \cite{lip}, Mo$_{3}$Al$_{2}$C \cite{mac}, Ru$_{7}$B$_{3}$ \cite{rb1,rb2}, Nb$_{0.18}$Re$_{0.82}$ \cite{nr1,nr2}, Re$_{6}$Zr \cite{rz1,rz2}, La$_{7}$Ir$_{3}$ \cite{li1}, have been investigated to understand the pairing mechanism and found that only few of them have exhibited unconventional superconducting behavior including triplet pairing \cite{tp, rz1, li1} and upper critical field close to the Pauli limit \cite{pl} others have shown only dominant s-wave characters.\\ 

In this paper, we present the synthesis, characterization and physical properties of intermetallic binary compound Re$_{6}$Hf, exhibiting bulk superconductivity at T$_{c}^{onset}$ $\sim$ 5.96 K. Re$_{6}$Hf crystallizes into cubic  non-centrosymmetric crystal structure with space group $I \bar{4}3m$. Low temperature magnetic, transport and heat capacity measurements suggest dominant s-wave character with upper critical field higher than the Pauli limiting field, indicating the signature of spin triplet component.\\

\section{EXPERIMENTAL DETAILS}

Polycrystalline sample of Re$_{6}$Hf (3g) was synthesized using standard arc melting technique where stoichiometric amount of $\mathrm{Hf(4N)}$ and $\mathrm{Re(4N)}$ taken in a nominal ratio of 6:1 on a water cooled copper hearth under the flow of high purity argon gas. Both the elements are melted together to make a single button of Re$_{6}$Hf, then flipped and remelted several times for the sample homogeneity. The sample formed is hard, silvery grey in color with negligible mass loss.\\
XRD was carried out in Panalytical diffractometer using Cu $K_{\alpha}$ radiation  ($\lambda$= 1.54056 $\text{\AA}$) for the characterization of crystal structure and phase purity.

Magnetization and AC susceptibility measurements were performed using Quantum Design Superconducting Quantum Interference Device (SQUID MPMS , Quantum Design) down to a temperature 1.8 K. In magnetization measurement the sample was cooled down to 1.8 K in zero field and then 10 mT field was applied. The measurement was carried out while warming it to 10 K called ZFCW mode whereas for FCC mode sample was cooled down in 10 mT field with data taken simultaneously. Magnetization and AC susceptibility measurements were also done in a temperature range between 1.8 K to 8.0 K under an applied magnetic fields up to 7.0 T.\\
The heat-capacity measurement was performed by the two tau time-relaxation method using the physical property measurement system (PPMS, Quantum Design, Inc.) in zero and applied magnetic fields up to 7.0 T.
The electrical resistivity measurements were performed in PPMS (Quantum Design, Inc.) by using conventional four-probe ac technique at 157 Hz and at excitation current of 10 mA. The measurements was done in zero field from 1.85 K to 300 K to know the resistivity $\rho_0$ and residual resistivity ratio (RRR).\\

\section{RESULTS AND DICUSSION}

\subsection{Sample characterization}
Polycrystalline sample of Re$_{6}$Hf was crushed into a very fine powder for XRD analysis. The powder XRD data collected from sample of Re$_{6}$Hf at room temperature shown in Fig.1 and were Rietveld refined by using High Score Plus Software. As evident from the pattern that sample was in single phase with no impurity traces found whatsoever. Crystal structure was confirmed to be cubic noncentrosymmetric $\alpha$-\textit{Mn} structure, primitive Bravais lattice $I \bar{4}3m$ (space group 217) with cubic cell parameter a = 9.6850(3)\text{\AA}. There are two formula units per unit cell. Refined cell parameters and atomic positions is shown in Table I

\begin{figure}
\includegraphics[width=9.cm,height=6.cm,origin=b]{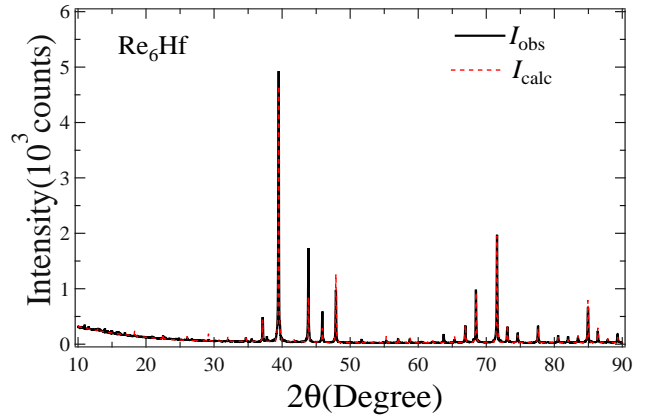}
\caption{\label{fig1:xrd} (color online) Powder XRD pattern for Re$_{6}$Hf sample recorded at room temperature using Cu $K_{\alpha}$ radiation. Rietveld refined calculated pattern for cubic $\alpha$-$\textit{Mn}$ (217) non-centrosymmetric structure shown by solid red line.}
\end{figure}

\begin{table}[h!]
\caption{Crystal structure Lattice and Rietveld refinement parameters obtained from the room temperature powder X-ray Diffraction of Re$_{6}$Hf}
Structure~~~~~~~Cubic\\
 ~~~~Space group~~~~~$I \bar{4}3m$\\
Lattice parameters\\
\textit{a}(\text{\AA})~~~~~~~~9.6850(3)\\
V$_{cell}$(\text{\AA}$^{3}$)~~~~~~~~908\\
\begin{center}
\begin{tabular}[b]{lcccc}\hline\hline
Atomic Coordinates\\
\hline
\\[0.5ex]
Atom &Wyckoff position & x & y & z\\                                  
Re1  &~~ 24g & 0.106 & 0.106 & 0.288\\             
Re2  &~~ 24g & 0.344 & 0.344 & 0.031\\
Hf1  &~~  8c & 0.315 & 0.315 & 0.315\\                       
Hf2  &~~  2a & 0 & 0 &  0\\
\\[0.5ex]
\hline\hline
\end{tabular}
\par\medskip\footnotesize
\end{center}
\end{table}

\subsection{Normal and Superconducting State properties}

\subsection{Electrical resistivity}
The normal state temperature dependence of the resistivity for Re$_{6}$Hf between 1.85 K and 300 K in zero magnetic field is shown in Fig.2(a), where the resistivity increases leisurely with temperature showing poor metallic behavior. The residual resistivity ratio is found to be $\rho(300/\rho(10)$ = 1.47, suggesting that the polycrystalline sample is dominated by disorder. This behavior is similar to other Re based $\alpha$-\textit{Mn} compounds  Nb$_{0.18}$Re$_{0.82}$ (RRR$\sim$1.3) \cite{nr1}, Re$_{24}$Ti$_{5}$ (RRR$\sim$1.3) \cite{rt} and Re$_{6}$Zr (RRR$\sim$1.10)\cite{rz1,rz2}. An expanded plot of the zero-field $\rho$(T) data is shown in the inset of Fig.2(a), which shows a superconducting transition with a transition width of $\Delta$T = 0.41 K, with an onset at T$_{c}^{onset}$ = 5.96 K and zero resistance at T = 5.55 K\\

The low temperature resistivity data has been fitted using power law 

\begin{equation}
\rho =\rho_0+AT^n
\label{eqn1:res}
\end{equation}

\begin{figure}
\includegraphics[width=9.cm,height=12.cm,origin=b]{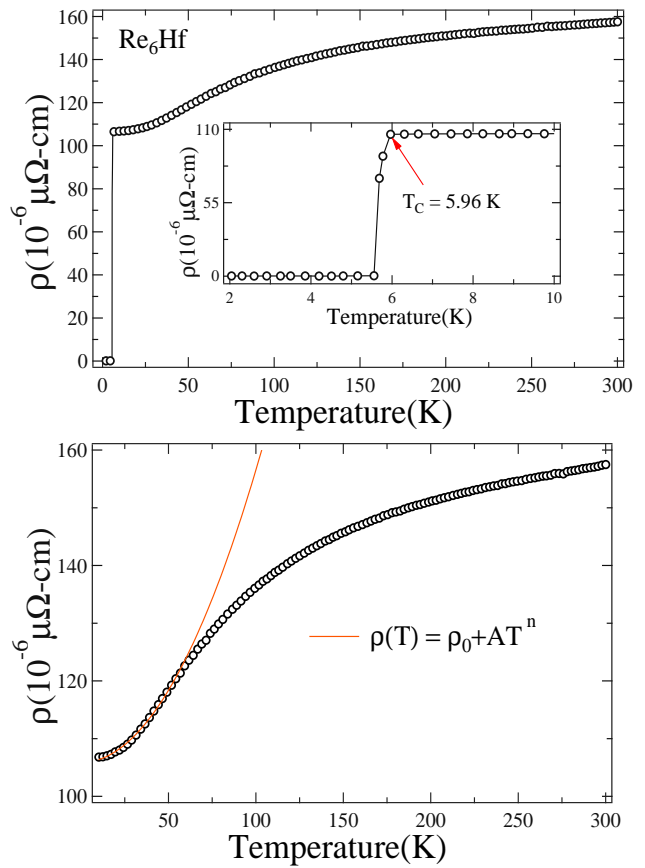}
\caption{\label{fig2:fl} (color online)(a) Temperature dependence of resistivity for Re$_{6}$Hf shown over the range 1.85 K $\le$ T $\le$ 300 K. Inset shows superconducting transition at T$_{c}^{onset}$ = 5.96 K. (b) Power law fitting over the temperature range 10 K $\le$ T $\le$ 50 K  gives n = 2.2 $\pm$ 0.1 indicating Fermi-liquid behavior at low temperature}
\end{figure}

from 10 K $\le$ T $\le$ 50 K, as shown in Fig.2(b) by the solid red line where $\rho_{0}$ is the residual resistivity due to crystallographic defects, A is the coefficient associated with the temperature dependence of electric resistivity which is taken as measure of degree of electron-electron correlation.

Using Eq. 1, we found  n $\sim$ 2.2 $\pm$ 0.1 with $\rho_{0}$ = 106.12 $\pm$ 0.02 $\mu\ohm$ cm and A = (2.34 $\pm$ 0.01) $\times$ 10$^{-3}$ $\mu\ohm$ cmK$^{2}$. The higher value of $\rho_{0}$ is most likely due to polycrystalline nature of the sample. The value of A is very low indicating a weakly correlated system whereas n is close to 2 suggesting a Fermi-liquid behavior in the normal state low temperature resistivity data.

\subsection{Magnetization}

The superconductivity in Re$_{6}$Hf was confirmed by the magnetization data taken in ZFCW and FCC mode as shown in Fig.3. The data was taken in low applied field of 10 mT, shows the onset of strong diamagnetic signal due to superconducting transition at T$_{c}^{onset}$ = 5.94 K. Superconducting transition was further confirmed by AC susceptibility (Fig.4) measurement.
\begin{figure}
\includegraphics[width=9.cm,height=6.cm,origin=b]{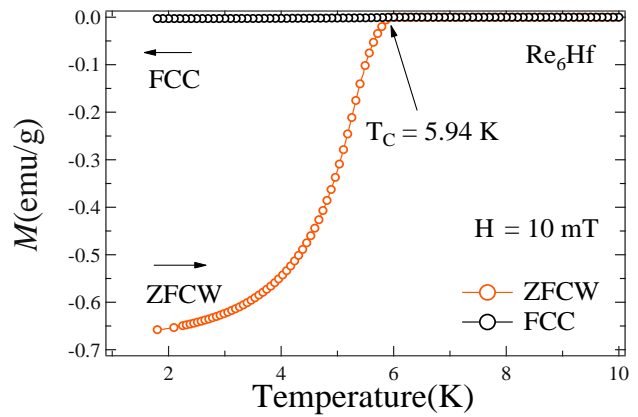}
\caption{\label{fig3:zfc} (color online) Temperature dependence of the magnetic moment, collected via zero-field cooled warming
(ZFCW) and field cooled cooling (FCC) methods under an applied field of 10 mT.}
\end{figure}

\begin{figure}
\includegraphics[width=9.cm,height=12.cm,origin=b]{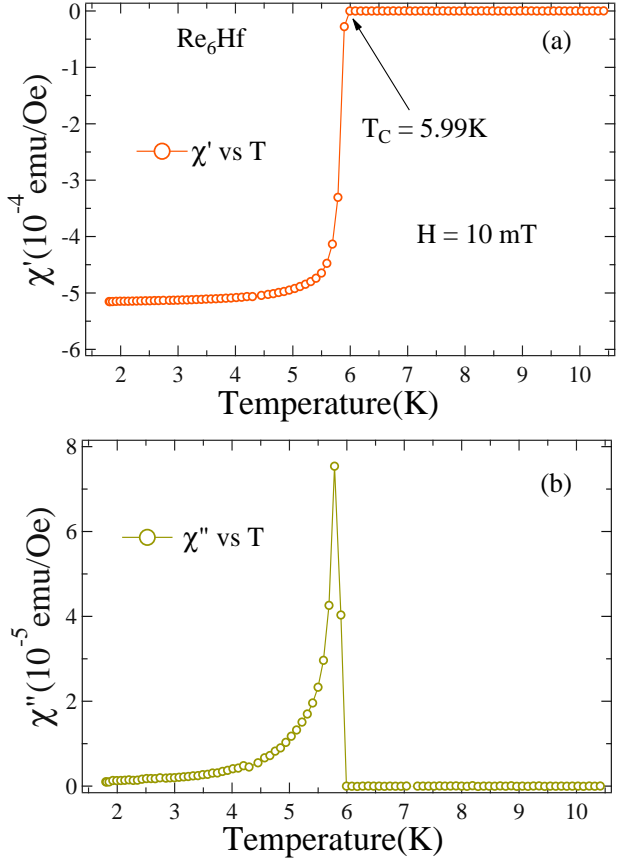}
\caption{\label{fig4:ac} (color online) Temperature dependence of AC susceptibility for Re$_{6}$Hf.}
\end{figure}
Fig.5(a) shows the low field (0-30 mT) magnetization M(H) curves for Re$_{6}$Hf taken at several temperatures. Lower critical field $H_{C1}(T)$ is defined as the field deviating from the linear line for the initial slope. The H$_{C1}$ is ~5.2 mT at T = 1.8 K and decreases monotonically with increase in temperature to ~1.0 mT at T = 5.5 K. The temperature variation of $H_{C1}(T)$ is shown in Fig.5(b), as expected $H_{C1}(T)$ is varying as a function of $T^{2}$ in accordance with Ginzburg-Landau (GL) theory

\begin{equation}
H_{C1}(T)=H_{C1}(0)\left(1-\left(\frac{T}{T_{c}}\right)^{2}\right)
\label{eqn2:HC1}
\end{equation} 

\begin{figure}
\includegraphics[width=9.cm,height=12.cm,origin=b]{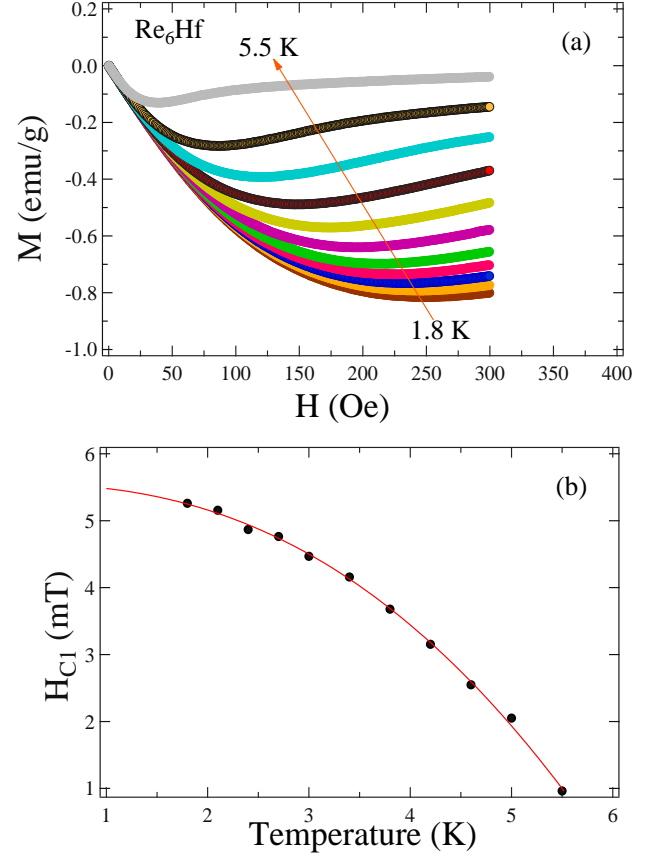}
\caption{\label{fig5:hc1} (color online) (a) Isothermal magnetization for Re$_{6}$Hf at various temperatures (b) Fitting using Eq. 2 gives  $H_{C1}(0)$ $\sim$ 5.6 mT}
\end{figure}
Lower critical field $H_{C1}(0)$ was estimated to be ~5.6 $\pm$ 0.5 mT by fitting the relation given in Eq. 2 in the data given in Fig.5(b).\\
A magnetization curve at 1.8 K in high applied magnetic field range ($\pm$ 7 T) is showed in Fig.6. As evident from the graph that Re$_{6}$Hf  exhibits a conventional type - II superconductivity. H$_{irr}$ was estimated from magnetization curve, is ~ 1.80 T at 1.8 K, therefore for fields H > H$_{irr}$ unpinning of vortices starts taking place.
          
\begin{figure}
\includegraphics[width=9.cm,height=6.cm,origin=b]{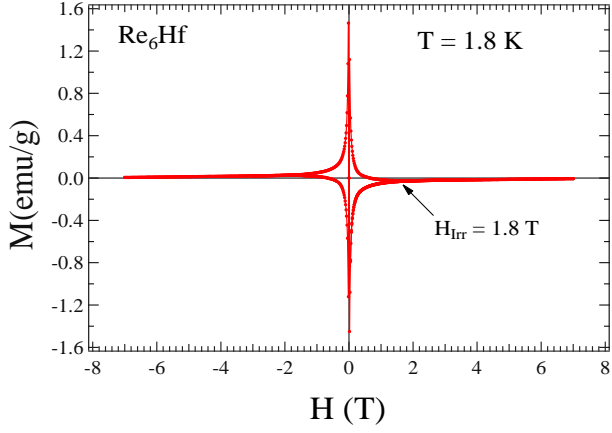}
\caption{\label{fig6:hys} (color online)  Magnetization vrs magnetic field variation at T = 1.8 K. The magnetization is irreversible below a field HIrr = 1.8 T.}
\end{figure}
                              
\subsection{Specific heat}

The specific heat $C(T)$ measurement for Re$_{6}$Hf in zero field in the temperature range 1.85 K $\le$ T $\le$ 10 K is shown in Fig.7(a). The superconducting transition in specific heat data is manifested by a sharp jump in $C(T)$ at $T_{c}$ = 5.96 K, confirming bulk superconductivity in Re$_{6}$Hf. The field dependence of the specific heat is shown in Fig.7(b). The specific heat jump near T$_{c}$ moves to lower temperature and the size of the jump $\Delta C$ becomes smaller with increasing magnetic field.
The normal state low temperature specific heat can be extracted easily with the relation 

\begin{equation}  
\frac{C}{T}=\gamma_{n}+\beta_{3}T^{2}+\beta_{5}T^{4}  
\label{eqn3:hc}    
\end{equation}                                                                                         
where the extrapolation of normal state behavior below T$_{c}$ to the T $\to$ 0 limit, allow the determination of normal state Sommerfeld coefficient $\gamma_{n}$, debye constant $\beta_{3}$ and anharmonic contribution $\beta_{5}$ to lattice heat capacity.
By fitting the above relation (3) between 36 K $\le$ T$^{2}$ $\le$ 100 K, shown by red line in Fig.8(a) we obtained $\gamma_{n}$=27.2 $\pm$ 0.1 mJ/mole$K^{2}$, $\beta_{3}$=0.28 $\pm$ 0.05 mJ/mole$K^{4}$ and $\beta_{5}$=0.0014 $\pm$ 0.0003 $\mu$J/mole$K^{6}$.

\begin{figure}
\includegraphics[width=9.cm,height=12.cm,origin=b]{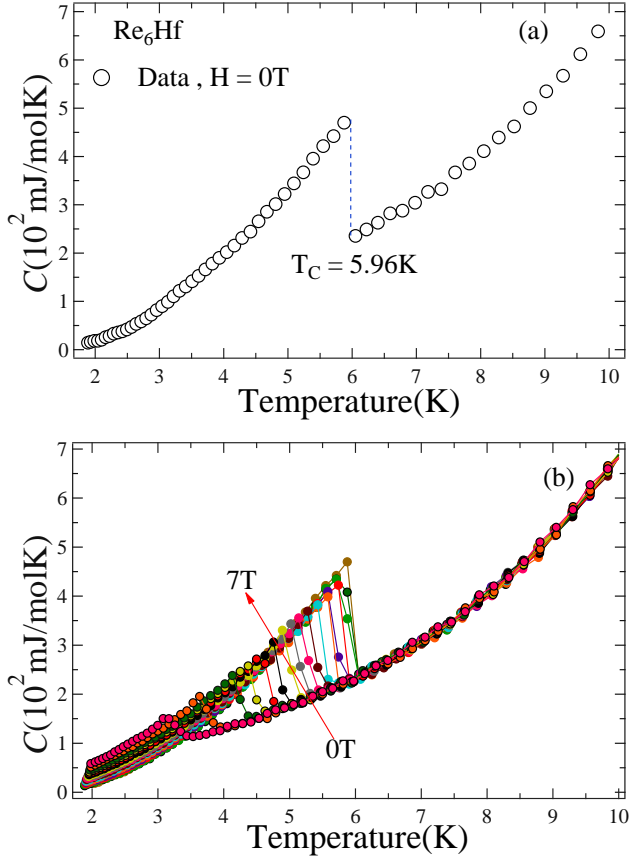}
\caption{\label{fig7:ct} (color online) (a) Temperature dependence of specific heat in zero-field showing superconducting transition at T$_{c}$ = 5.96 K (b) Field dependence of the specific heat data,  showing the suppression of T$_{c}$ with increasing field.}
\end{figure}

The Debye temperature $\theta_{D}$ is related to $\beta_{3}$ coefficient through formula
\begin{equation} 
\theta_{D}= \left(\frac{12\pi^{4}RN}{5\beta}\right)^{\frac{1}{3}}
\label{eqn4:dt}  
\end{equation}
                          
where R is the molar gas constant(=8.314 $Jmole^{-1} K^{-1}$). Using N (= 14) the number of atoms per unit cell, gives $\theta_{D}$ = 365 K. For non-interacting particles the Sommerfeld coefficient is proportional to density of states $D_{C}(E_{F})$ at the Fermi level which is calculated to be 11.53 $\frac{states}{eVf.u}$ from the relation given by
\begin{equation} 
\gamma_{n}= \left(\frac{\pi^{2}k_{B}^{2}}{5}\right)D_{C}(E_{f})
\label{eqn5:ds}  
\end{equation}  
where k$_{B}$$\approx$ 1.38 $\times$ 10$^{-23}$ J K$^{-1}$. \\                            
This  calculated  density of states $D_{C}(E_{f})$ and the effective mass $m^{*}$ of quasi-particles contains the influence of many body electron-phonon interaction and is related to bare band-structure density of states $D_{band}(E_{f})$ and $m^{*}_{band}$  according to 
\begin{equation}
D_{C}(E_{f}) = D_{band}(E_{f})(1+\lambda_{e-ph})\\ 
\label{eqn6:dk}                                                                        
\end{equation}

\begin{equation}
m^{*} = m^{*}_{band}(1+\lambda_{e-ph})\\
\label{eqn7:mel}                                                                        
\end{equation}
where $\lambda_{e-ph}$ is the  dimensionless electron phonon coupling constant.
Electron-phonon coupling constant which gives the strength of attractive interaction between electron-phonon can be calculated using  $\theta_{D}$  and  $T_{C}$ given in McMillan theory \cite{Mc} by
\begin{equation}
\lambda_{e-ph} = \frac{1.04+\mu^{*}ln(\theta_{D}/1.45T_{c})}{(1-0.62\mu^{*})ln(\theta_{D}/1.45T_{c})-1.04 }
\label{eqn8:ld}
\end{equation}                       
where $\mu^{*}$ is the repulsive screened coulomb parameter, typically given by $\mu^{*}$ = 0.13 for many intermetallic superconductors. Using $T_{c}$= 5.96 K (from our specific heat measurement) we obtained $\lambda_{e-ph}$ = 0.63. This value is comparable to other fully gapped non-centrosymmetric superconductors such as 0.6 for Re$_{24}$Ti$_{5}$ \cite{rt}, 0.5 for Nb$_{0.18}$Re$_{0.82}$ \cite{nr1}, Re$_{6}$Zr \cite{rz2}, SrAuSi$_{3}$ \cite{sas} suggesting that Re$_{6}$Hf is a moderately coupled superconductor.
Using the value of $\lambda_{e-ph}$, we calculated $D_{band}(E_{f})$ to be equal to 7.04 $\frac{states}{eVf.u}$, also by  taking $m^{*}_{band}= m_{e}$ the effective mass for the quasi-particles comes out be 1.63 $m_{e}$.
The mean free path $\textit{l}$ can be estimated using the relation
\begin{equation} 
\textit{l} = 3\pi^{2}\left(\frac{\hbar}{e^{2}\rho_{0}} \right)\left(\frac{\hbar}{m^{*}v_{f}}\right)^{2}
\label{eqn9:l} 
\end{equation}                                                                                                               
where the Fermi velocity v$_{f}$ is given by
\begin{equation}
v_{f} = \frac{\pi^{2}\hbar^{3} D_{C}(E_{f})}{ m^{*2} V_{f.u}} 
\label{eqn10:fv}   
\end{equation}                                                                                
where V$_{f.u}$ = $\frac{V_{cell}}{2}$ is the volume per formula unit. Using D$_{C}$(E$_{f}$) value 11.53 $\frac{states}{eVf.u}$ we calculated v$_{f}$ = 8.27 $\times$ 10$^{5}$ cm/s which gives mean free path $\textit{l}$ = 8.5 \text{\AA}

The electronic contribution to specific heat in superconducting state $C_{el}(T)$ for  Re$_{6}$Hf is shown in Fig.8(b) and can be calculated by subtracting the phononic contribution from the measured data $C(T)$
\begin{equation}
C_{el}(T) = C(T)-\beta^{3}T^{3}+\beta^{5}T^{5}\\
\label{eqn11:cel}
\end{equation} 

The magnitude of specific heat jump $\frac{\Delta C_{el}}{T_{c}}$ at T$_{c}$ in C$_{el}$/T data is 41.55 mJ/moleK$^{2}$ . This gives the normalized specific heat jump $\frac{\Delta C_{el}}{\gamma_{n}T_{c}}$ = 1.53 for $\gamma_{n}$ =27.2 mJ/moleK$^{2}$ which is slightly higher than the BCS value of $\frac{\Delta C_{el}}{\gamma_{n}T_{c}}$ = 1.43 in the weak coupling limit.

The temperature dependence of specific data in the superconducting state can best be described by an exponential function given by C$_{el}$/T = aexp$\left(\frac{-bT_{c}}{T}\right)$ where b = $\Delta(0)$/k$_{B}$T$_{c}$ for fully gaped s - wave superconductor. As shown in Fig.8(b) we have reasonably good agreement with the above function to our experimental data C$_{el}$/T for b=$\alpha$ = 1.75, which is close to value for a BCS superconductor b = $\Delta(0)$/k$_{B}$T$_{c}$ = 1.764. Therefore the above analysis suggests that for Re$_{6}$Hf  can simply accounted for a BCS type fully gaped superconductor. 
\begin{figure}
\includegraphics[width=9.cm,height=12.cm,origin=b]{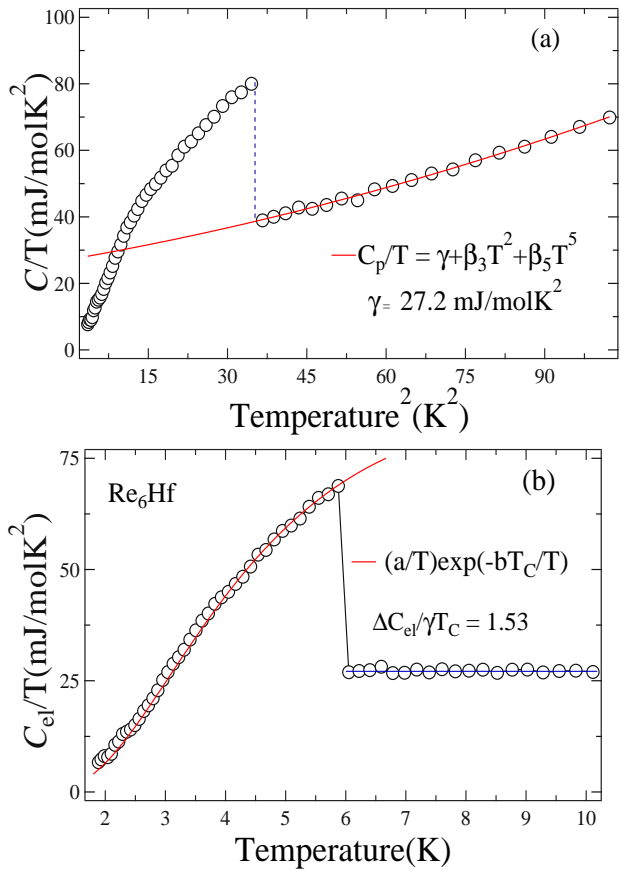}
\caption{\label{fig8:exp1} (color online) (a) $C$/T vs T$^{2}$ for 36 K $\le$ T$^{2}$ $\le$ 100 K in zero field well described by the equation $\frac{C}{T}=\gamma_{n}+\beta_{3}T^{2}+\beta_{5}T^{4}$ (b) Solid red curve in C$_{el}$/T versus T data shows exponential fitting for b = $\alpha$ = 1.75 which is close  the BCS value is $\alpha$ =  1.764.}
\end{figure}

The fully gaped s-wave superconductivity in Re$_{6}$Hf can also be confirmed by the magnetic field dependence of the Sommerfeld coefficient $\gamma(H)$. The Sommerfeld coefficient is proportional to the quasiparticle density of states, hence as we apply more field, the vortex density increases due to an increase in the number of field induced vortices which in turn enhances the quasiparticle density of states. This gives rise to linear relation between $\gamma$ and H i.e., $\gamma$(H) $\propto$ H for a nodeless and isotropic s- wave superconductor \cite{cc,pg,sas}.
For a superconductor with nodes in gap Volovik predicted a non -linear relation given by $\gamma$(H) $\propto$ H$^{\frac{1}{2}}$ \cite{gev} 

Field dependence of $\gamma$ is shown in Fig.9 where Sommerfeld coefficient $\gamma$ was calculated by fitting the relation (12) in C$_{el}$/T versus T data for various fields and extrapolating it to T = 0 K \cite{sas}

\begin{equation}
\frac{C_{el}}{T} = \gamma + \frac{A}{T}exp\left(\frac{-bT_{c}}{T}\right)
\label{eqn12:gh}
\end{equation}
here we can clearly observe a linear relation between $\gamma$(H) and H confirming the s - wave superconductivity in Re$_{6}$Hf which we have already seen by the exponential signature in at low temperature specific heat measurement.

\begin{figure}
\includegraphics[width=9.cm,height=6.cm,origin=b]{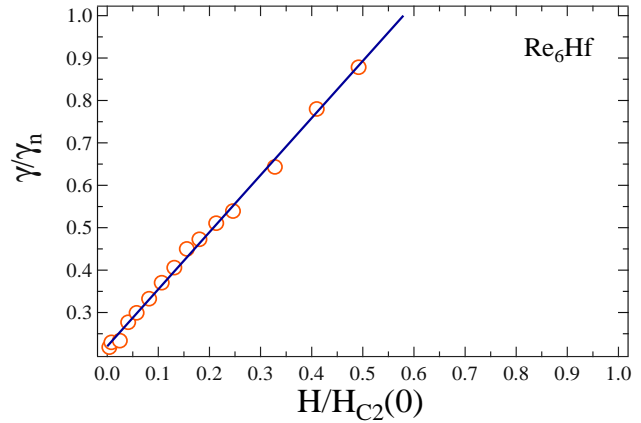}
\caption{\label{fig9:gh} (color online) $\gamma$ and H was plotted versus each other after normalizing it by $\gamma_{n}$(=27.2 mJ/molK$^{2})$ and H$_{C2}$(0)(=12.2 T). Solid red curve in $\gamma/\gamma_{n}$ versus H/H$_{C2}$(0) shows the linear relation indicating s - wave superconductivity in Re$_{6}$Hf}
\end{figure}

The condensation energy U (0) can be calculated using $\gamma$ and $\alpha(=\Delta (0)/k_{B}T_{c}$ from the relation 
\begin{equation}
 U(0) = \frac{1}{2}\Delta^{2}(0)D_{band}(E_{f}) = \frac{3\gamma_{n}\Delta^{2}(0)}{4\pi^{2}k_{B}^2}
\label{eqn13:ce}
\end{equation}                                                              
which gives U (0) = 225.4 mJ/mole.
The magnitude of electron - electron correlation can be given by Kadowaki- Woods ratio which is given by K$_{w}$ =$\frac{A}{\gamma_{n}^2}$, where A is the coefficient of the quadratic temperature dependent resistivity term, whereas $\gamma_{n}$ Sommerfeld coefficient . In heavy fermionic systems where electron- electron correlation is significant Kadowaki- Woods ratio K$_{w}$ approaches 1 $\times$ 10$^{-5}$ $\mu\ohm$cm mJ$^{-2}$mole$^{-2}$K$^{2}$ \cite{kw1,kw2,Sc}. For A=0.0023 $\mu\ohm$ cm K$^{2}$  and  $\gamma_{n}$=27.2 mJ/mole K$^{2}$, we found K$_{w}$= $0.31 \times 10^{-5}$ $\mu\ohm$cm mJ$^{-2}$mole$^{-2}$K$^{2}$  which is smaller than strongly correlated systems indicating that Re$_{6}$Hf is a weakly correlated system.

Isothermal magnetization curves are measured in magnetic fields upto 7.0 T to calculate the upper critical field H$_{C2}$(0) as showed in inset of Fig.10. The transition temperature was taken as the onset of diamagnetic signal in magnetization measurement. As the field is increased T$_{C}$ shifts to lower temperature with superconducting transition becoming broader. H$_{C2}$(T) was seen to be vary as linearly when plotted against reduced temperature(T/T$_{c}$) as displayed in Fig. 10, and can be fitted using Ginzburg-Landau(GL) formula to calculate H$_{C2}$(0)

\begin{equation}
H_{C2}(T) = H_{C2}(0)\frac{(1-t^{2})}{(1+t^2)}
\label{eqn14:hc2}
\end{equation}

where t=T/T$_{c}$. Our experimental data fits fairly well with equation (14) yielding the upper critical field H$_{C2}$(0) to be around 12.2 $\pm$ 0.1 T. Similar linear behavior of H$_{C2}$(T) was noted when calculated from specific heat measurements $C$(T) as shown by blue data points in Fig. 10, when fitted with GL formula gives H$_{C2}$(0) $\approx$ 12.2 T \\

\begin{figure}
\includegraphics[width=9.cm,height=6.cm,origin=b]{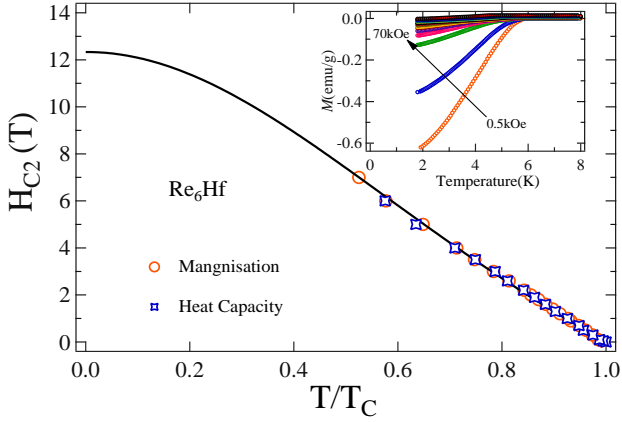}
\caption{\label{fig10:Mtdif} (color online) (a)  Determination of the upper critical field via magnetization and heat capacity measurements.The solid lines are Ginzburg Landau (Eq. 14) fits. Inset shows the $\textit{M(T)}$ curves for various applied magnetic field }
\end{figure}

H$_{C2}$(0) can be used to estimate the Ginzburg Landau coherence length $\xi_{GL}$ from the relation \cite{tin} 
\begin{equation}
H_{C2}(0) = \frac{\Phi_{0}}{2\pi\xi_{GL}^{2}}
\label{eqn15:up}
\end{equation}   
                                                                                   
where and $\Phi_{0}$ (=2.07 $\times$ 10$^{-15}$ Tm$^{2}$) is the magnetic flux quantum. Using H$_{C2}(0)$ = 12.2 T we calculated $\xi_{GL}(0)$ $\approx$ 52 \text{\AA}, mean free path $\textit{l}$ which is calculated from normal state specific heat data was 8.5 \text{\AA} is much lower than $\xi_{GL}(0)$ indicating the dirty limit superconductivity in Re$_{6}$Hf.

In a type-II superconductor the Cooper pair breaking due to applied magnetic field is attributed to two types of mechanisms Orbital and Pauli paramagnetic limiting field effect. In orbital pair breaking field-induced kinetic energy of Cooper pair exceeds the superconducting condensation energy  whereas in Pauli paramagnetic limiting it is energetically favorable for the electron spins to align with magnetic field, thus breaking the Cooper pairs. 
For BCS superconductors the orbital limit of upper critical field  H$_{C2}^{orbital}$(0) is given by using the WHH expression \cite{EH,NRW}
\begin{equation}
H_{C2}^{orbital}(0) = -\alpha T_{c}\left.\frac{-dH_{C2}(T)}{dT}\right|_{T=T_{c}}
\label{eqn16:whh}
\end{equation}
  
where $\alpha$ is the purity factor given by 0.693 for dirty limit superconductors and 0.73 for clean limit superconductors.     Initial slope $\left.\frac{-dH_{C2}(T)}{dT}\right|_{T=T_{c}}$ to be 2.36 T/K from H$_{C2}$-T phase diagram obtained earlier which gives us the orbital limiting upper critical field  H$_{C2}^{orbital}$(0) in dirty limit 9.77 T.
The Pauli limiting field within the BCS theory is given by $H_{C2}^{p}$(0)= 1.86 T$_{c}$\cite{BC,AM}, which for T$_{c}$ = 5.96 K gives $H_{C2}^{p}$(0)$\approx$ 11.10 T.
The Maki parameter \cite{MK} $\alpha_{M}$ is a measure  relative  strengths of the orbital and Pauli- limiting values of H$_{C2}$ and is given by
\begin{equation}
\alpha_{M} = \sqrt{2}H_{C2}^{orb}(0)/H_{C2}^{p}(0) 
\label{eqn17:up}
\end{equation}                                   
from above relation we obtain $\alpha_{M}$ = 1.24 .The sizable value of Maki parameter obtained from this expression is an indication that effect of Pauli limiting field is non-negligible \cite{nr2,rz1}.

The Ginzburg Landau penetration depth $\lambda_{GL}$(0) can be obtained from the H$_{C1}$(0) and $\xi_{GL}(0)$ using the relation \cite{tkf}
 
\begin{equation}
H_{C1}(0) = \left(\frac{\Phi_{0}}{4\pi\lambda_{GL}^2(0)}\right)ln\left(\frac{\lambda_{GL}(0)}{\xi_{GL}(0)}\right)  
\label{eqn18:ld}
\end{equation}                                                                     
For H$_{C1}$(0) $\approx$ 5.6 mT and $\xi_{GL}(0)$ = 52 \text{\AA}, we obtain $\lambda_{GL}(0) \approx 3538 \text{\AA}$ 
The Ginzburg Landau parameter is given by the relation 
\begin{equation}
k_{GL} = \frac{\lambda_{GL}(0)}{\xi_{GL}(0)}
\label{eqn19:kgl}
\end{equation}
For  $\xi_{GL}(0)$ = 52  \text{\AA} and $\lambda_{GL}(0)$ = 3538  \text{\AA} gives k$_{GL}$ $\approx$ 68. This value of $k_{GL} \gg \frac{1}{\sqrt{2}}$ indicating that Re$_{6}$Hf is a strong type - II superconductor as already seen from magnetization curve in Fig. 6.
Thermodynamic critical field H$_{C}$ is be obtained from $k_{GL}$ , H$_{C1}$(0) and  H$_{C2}$(0) using the relation \cite{tkf} 
\begin{equation}
H_{C1}(0)H_{C2}(0) = H_{C}^2lnk_{GL}
\label{eqn20:tf}
\end{equation}                                                                                 
yielding H$_{C}$ around 126.7 mT.

The Ginzburg number G$_{i}$ which is basically the ratio of thermal energy $K_{B}$T (with upper limit of T is T$_{c}$) to condensation energy associated with coherence volume measure the sensitivity of vortex system against thermal fluctuations is obtained by the from the relation     \cite{vortex,omi}

\begin{equation}
 G_{i} = \frac{1}{2}\left(\frac{k_{B}\mu_{0}\tau T_{c}}{4\pi\xi^{3}(0)H_{C2}(0)}\right)^2
\label{eqn21:gi}
\end{equation}   
                                                                        
where $\tau$ is anisotropy parameter which is 1 for cubic Re$_{6}$Hf. For $\xi_{0}$ = 52  \text{\AA}, H$_{C}$(0) = 126.7 mT and T$_{c}$= 5.96 K we got G$_{i}$= 6.65 $\times$ 10$^{-2}$. As it was already noticed from magnetization curve that melting of vortices starts at field H > H$_{irr}$(=1.8 T) which is much smaller than upper critical field at 1.8 K H$_{C2}$(1.8 K)= 10.16 T. This behavior usually found in high temperature superconductors where melting of vortices attributed to thermal fluctuations and very rare in low T$_{c}$ superconductors. G$_{i}$ which is found to be of the order same as high T$_{c}$ superconductors, suggests that thermal fluctuations may be playing important role in unpinning of vortices. \\

\begin{table}[h!]
\caption{Normal and superconducting properties of Re$_{6}$Hf}
\begin{center}
\begin{tabular}[b]{lcc}\hline\hline
Atomic Coordinates& unit& value\\
\hline
\\[0.5ex]                                  
T$_{c}$& K& 5.96\\             
$\rho_{0}$& $\mu\ohm$cm& 106.12 \\
A& $\mu\ohm$cmK$^{-2}$& 0.0023\\
H$_{C1}(0)$& mT& 5.6 \\                       
H$_{C2}(0)$& T& 12.2 \\
H$_{C}(0)$& mT& 126.7 \\
H$_{C2}^{orbital}(0)$& T& 9.8\\
H$_{C2}^{P}(0)$& T& 11.1\\
$\xi_{GL}$& \text{\AA}& 52\\
$\lambda_{GL}$& \text{\AA}& 3538\\
$k_{GL}$& &68.0\\
$\textit{l}$& \text{\AA}& 8.5\\
$\gamma$& mJmol$^{-1}$K$^{-2}$& 27.2\\
$\beta$ & mJmol$^{-1}$K$^{-4}$& 0.28\\
$\theta_{D}$& K& 365\\
$\lambda_{ep}$&  &0.63\\
D$_{C}$(E$_{f}$)& states/ev f.u& 11.53\\
$\Delta C_{el}/\gamma_{n}T_{C}$&   &1.53\\
$\Delta/k_{B}T_{C}$&~~  &1.75\\
\\[0.5ex]
\hline\hline
\end{tabular}
\par\medskip\footnotesize
\end{center}
\end{table} 
\section{Conclusion}
Single phase polycrystalline sample of Re$_{6}$Hf was prepared by arc- melting technique, which crystallizes into non-centrosymmetric $\alpha$-$\textit{Mn}$ (217) structure with lattice parameter $\textit{a}$ = 9.6850(3) \text{\AA}. Transport, magnetization and thermodynamic measurements reveals that Re$_{6}$Hf is a weakly correlated type-II superconductor with bulk transition at T$_{c}$ $\approx$ 5.96 K. Above transition temperature T$_{C}$, resistivity measurement shows poor metallic behavior whereas magnetization measurement showed weak paramagnetism. Mean free path calculated $\textit{l}$ (=8.5 \text{\AA}) is much smaller than coherence length $\xi$ (= 52 \text{\AA}) confirms dirty limit superconductivity in Re$_{6}$Hf. Low temperature specific heat measurement in superconducting regime indicates that Re$_{6}$Hf is a moderately coupled superconductor $\lambda_{e-ph}$ = 0.63 with normalized specific heat jump  $\frac{\Delta C_{el}}{\gamma_{n}T_{c}}$ = 1.53. Computed value of upper critical field H$_{C2}$ is greater than Pauli limiting field indicates that there may be spin triplet component in superconducting condensate, however specific heat data in superconducting state fits perfectly well with thermally activated behavior manifested by exponential relation C$_{es}$ $\propto$ exp$\left(\frac{-bT_{c}}{T}\right)$ indicating that Re$_{6}$Hf is a s-wave superconductor. BCS type superconductivity in Re$_{6}$Hf was further confirmed by field dependence of Sommerfeld coefficient $\gamma(H)$ where it varies linearly with H suggesting that electronic states near the Fermi surface are fully gaped. Similar type of behavior observed in Re$_{6}$Zr \cite {rz1} by $\mu$SR spectroscopy. Our initial $\mu$SR measurements suggest the presence of triplet component which is coherent with current result.

\section{Acknowledgments}

R.~P.~S.\ acknowledges Science and Engineering Research Board, Government of India for the Ramanujan Fellowship.

\end{document}